\documentclass[12pt]{article}
\usepackage{color}
\usepackage[usenames,dvipsnames,svgnames,table]{xcolor}
\definecolor{dark-red}{rgb}{0.7,0.1,0.1} 
\definecolor{dark-blue}{rgb}{0,0,0.7} 
\usepackage[linkcolor=dark-red,
            colorlinks=true,
            urlcolor=dark-blue,
            pdfstartview={XYZ null null 1.00},
            pdfauthor={Gaurav Sood, gsood07@gmail.com},
            citecolor=dark-red,
            bookmarks=false,
            pdfborder={0 0 0},
            pdftitle={Fairly Random}]{hyperref}
            
\usepackage{amsfonts,amssymb,amsbsy,amsmath,amsxtra}

\usepackage[letterspace=1000]{microtype}
\usepackage{libertine}
\usepackage[T1]{fontenc}

\usepackage{indentfirst}
\usepackage{setspace} 
\usepackage{multirow}

\usepackage{verbatim}

\usepackage[multiple]{footmisc}
\usepackage{fancyvrb}

\usepackage{longtable}

\usepackage[margin=1in]{geometry}
\usepackage{graphicx}

\raggedright
\usepackage[authoryear]{natbib}
\usepackage{url}

\begin{document}
\title{\vspace{-.5cm}\normalsize{Fairly Random: The Impact of Winning the Toss on the Probability of Winning\footnote{The paper benefitted from comments by Andrew Gelman, Donald Green, and Daniel Stone. Scripts used to scrape the data, the final data set, and scripts used for analysis can be found at: \href{https://github.com/dwillis/toss-up}{https://github.com/dwillis/toss-up}. }\vspace{.5cm}}}
\author{\normalsize{Gaurav Sood}\footnote{Gaurav is an independent researcher. He can be reached at: \href{mailto:gsood07@gmail.com}{\small{gsood07@gmail.com}}} \and \normalsize{Derek Willis}\footnote{Derek is a  news applications developer at ProPublica. He can be reached at: \href{mailto:dwillis@gmail.com}{\small{dwillis@gmail.com}}}}
\date{\vspace{.5cm}\normalsize{\today}}
\maketitle

\begin{center}\textbf{Abstract}\\\end{center}
In a competitive sport, every little thing matters. Yet, many sports leave some large levers out of the reach of the teams, and in the hands of fate. In cricket, world's second most popular sport by some measures, \footnote{See \citet{Economist2011} in The Economist.} one such lever---the toss---has been subject to much recent attention. Using a large novel dataset of 44,224 cricket matches, we estimate the impact of winning the toss on the probability of winning. The data suggest that winning the toss increases the chance of winning by a small ($\sim$ 2.8\%) but significant margin. The advantage varies heftily and systematically, by how closely matched the competing teams are, and by playing conditions---tautologically, winning the toss in conditions where the toss grants a greater advantage, for e.g., in day and night matches, has a larger impact on the probability of winning.  
\clearpage

In nearly all cricket matches, it is claimed that there is a clear advantage to either bowling or batting first. The advantage is pointed to by commentators, by the team captains in the pre-toss interview, and by the captain of the losing team in the post-match interview. And to wrest the said advantage, the captain merely needs to pick the side of the coin that will be left facing the sky after the toss. And while this method of granting advantage is fair on average, the system isn't fair in any one game. 

At first glance, this imbalance seems inevitable. After all, someone has to bat first. One can, however, devise a baseball like system, with a series of short innings woven together. If that violates the nature of the game too much, one can easily create pitches that don't deteriorate appreciably over the course of a match. Or, one can come up with an estimate of the advantage, and adjust the scores accordingly, akin to an adjustment that is issued when the matches are shortened due to rain.

None of this to say that there is actually an advantage in winning the toss, or that teams are able to successfully exploit any such advantage. For it may be impossible to predict well in advance the advantage of bowling or batting first. (If pre-match assessments of the pitch by media commentators are anything to go by, error in assessment of conditions is likely large.) Or, it may be that teams squander the potential advantage by using bad heuristics to choose what they do. For instance, teams may weigh outcomes from recent matches `too' heavily; e.g., a team that has, of late, won chasing may choose to chase even though pitch conditions favor batting first.

To assess the net observed advantage of winning the toss, we exploit data from a novel dataset of over 43,000 first-class men's cricket matches--- to our knowledge, the largest ever dataset assembled for the question, and nearly 50--100 times larger than used in prominent previous attempts \citep[see,][]{dawson2009bat, de1998winning}. In analyzing these data, we avoid a common but important pitfall that some other studies on the topic fall into. To avoid post-treatment bias \citep[see][]{acharya2015}, unlike \citet{dawson2009bat}, \citet{Saad2015}, etc., we do not condition on post coin-toss decisions. We find that winning the toss grants a small but significant advantage, but that advantage varies considerably and systematically. We next assess whether the advantage of winning the toss varies, broadly speaking, by how closely matched the teams are, and by how large an advantage winning the toss grants---the advantage of winning the toss is greater in certain playing conditions than others. We find that the advantage of winning the toss varies widely and systematically, in expected ways.

\section*{Data}
Data are from 44,224 first-class cricket matches. It is a near census of the relevant population.\footnote{Data excludes scheduled matches that were abandoned without the toss being conducted.} We have data on all types of matches: domestic and international Twenty20s---T20s and T20Is respectively, domestic and international one-dayers---List A and One-Day Internationals (ODI) respectively, and domestic and international multi-day matches --- First Class (FC) and Tests respectively.\footnote{There is a rich variety of first-class matches. In English county cricket, first-class matches last four days. Some first class matches last just a day. Others two days. Yet others three days. And till a particular point in history, a test match lasted as long as it was needed to finish a game. We elide over such differences.} 

Of the 44,224 matches, 1,019 matches were abandoned without play. We exclude these matches. In another 1,376 matches, we do not have information on whether the team chose to bat or bowl after the toss. Informal inspection suggests that data are missing because no match was played. We excludes these matches as well. 

In limited overs cricket, a minimum number of overs must be bowled to establish a result. In a one-day match, for instance, each side must bat at least 20 overs for a result to be declared. In 769 matches, or roughly 1.7\% of the total matches, not enough overs were bowled to get a result. We exclude these matches from our analysis. This leaves us with data from 41,060 matches. We analyze these data.
 
\section*{Analyses and Results}

We assume that the outcome of a toss is random. Conditional on the outcome of a toss being random, the effect of winning the toss can be attributed to the toss itself. Any decision made after the outcome of the toss is known, however, is `post-treatment.' In particular, the decision to bat or bowl first is made after accounting for the relative strengths and weaknesses vis-\`{a}-vis the competing team at that particular instance, and thus not independent of team attributes. Hence, conditioning on decision to bowl or bat first can bias estimates of advantage of winning a toss. Thus, unlike \citet{dawson2009bat}, \citet{Saad2015}, we solely rely on the assumption that the outcome of a toss is random. 

But before we exploit the design, we shed some light on the validity of the assumption. In particular, we assess whether the coin toss is somehow rigged, with the home side enjoying the rub of the green more often. For this analysis, we only get to exploit international matches as establishing which of the teams is the home team in local matches is somewhat arduous. Of the 5,684 international matches for which we can match the country of the ground to the country of one of the teams, the home team won the toss in 2,892 matches, or about 50.87\%. The chance of getting as many wins by fluke after tossing 5,684 coins is about 18.91\%. The chance is low, but not eyebrow raisingly so. 

However, rather than consider all home matches, we may instead want to only consider matches that are officiated only by home umpires---the norm till 1992.\footnote{For more information on move to neutral umpires, see \href{http://www.espncricinfo.com/magazine/content/story/511175.html}{Neutral Umpires} by S. Rajesh on ESPNcricinfo.} In matches featuring umpires from only the home country, the team with `home umpire' advantage won toss nearly 51.9\% of the times. And the chance of getting a greater percentage of wins than 51.9\% in 2,965 is a shade less than 4\%. Thus, there is some reason to worry that the tosses are rigged. Any such rigging would bias estimates of advantage of winning the coin toss to the extent that it is correlated with ability. More plainly, if stronger teams win more tosses, estimates of the advantage of the coin toss would be inflated upwards. And vice versa, if otherwise. We, however, do not have good reasons to think that there is a correlation. So for now, we proceed as if the tosses are random. 

Another caveat about interpretation before we present the results. As we discuss in the introduction, we cannot estimate the actual advantage of winning a toss. We can only estimate the net observed advantage, which is the extent to which the teams capitalize on the potential advantage. With that, the results.

The team that wins the toss wins the match 2.8\% more often than the team that loses the toss. This is a reasonable sized advantage in a competitive sport --- though likely much smaller than the number that most commentators carry in their heads. This advantage, however, varies by format, by conditions, by whether or not a particular formula was used to adjust scores when it rains, and how much better the team that won the toss is vis-\`{a}-vis the competing team. Much of the variability follows expected patterns.

\begin{figure}[htbp]
\caption{Difference in Winning Percentages of Teams that Won the Toss and Teams that Lost the Toss by Type of Match.}
\centering
\includegraphics[scale=1]{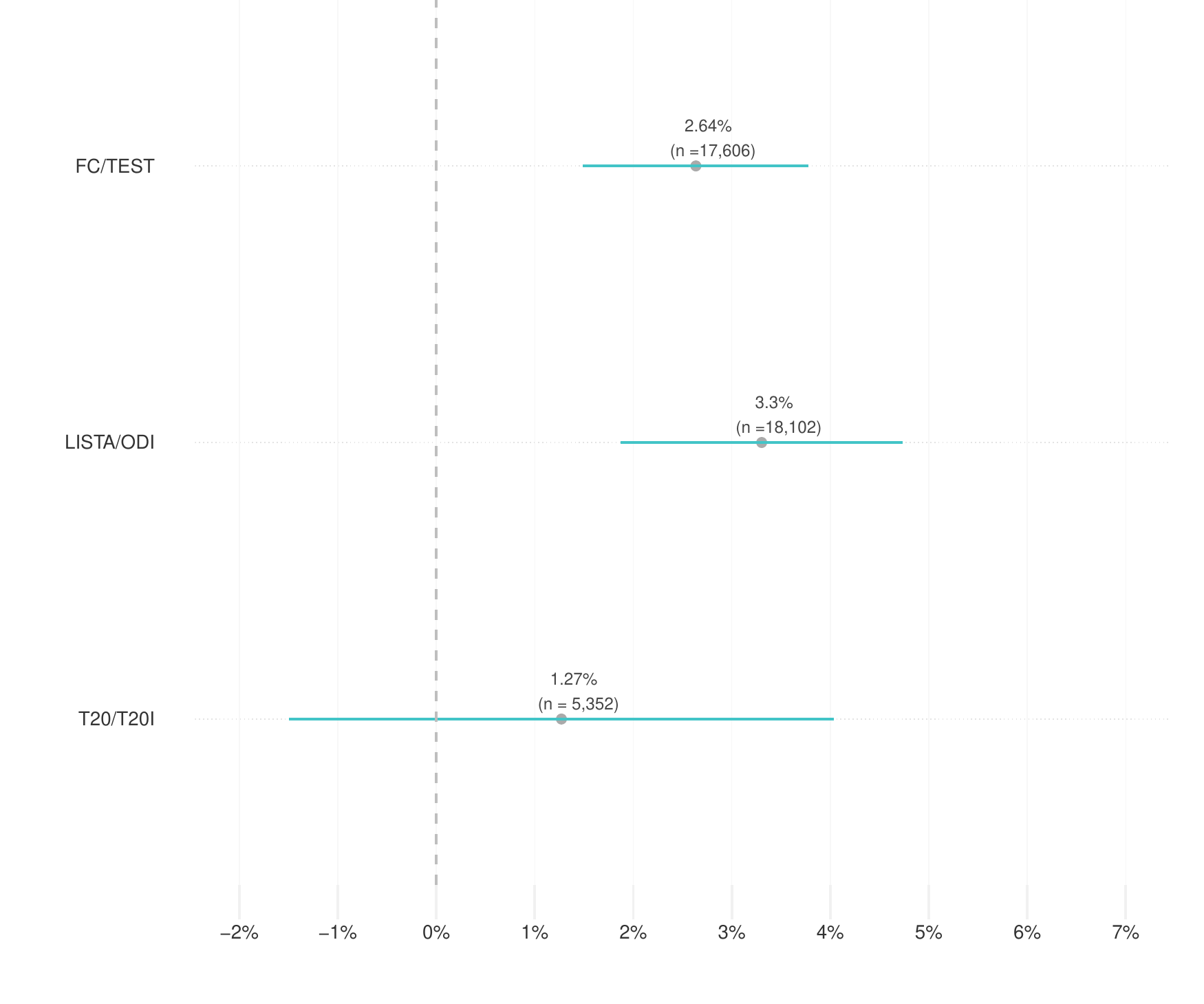}
{\footnotesize \\ Note: Means and 95\% confidence intervals. \emph{n} refers to the number of matches.\par}
\label{fig:type}
\end{figure}

The conventional wisdom among lay cricket followers is that toss grants the greatest advantage in multi-day affairs like tests and first class matches, followed by day long affairs, and Twenty20s. And there is good dose of common sense behind the conventional wisdom. Pitches invariably deteriorate over multiple days and batting last in a test match is often the most challenging time to bat. The pitch deteriorates far less over the course of the day, or in case of Twenty20s, a few hours. And indeed unlimited over matches provide the greatest advantage--- the average advantage over FC and test matches is north of 2.6\% (see Figure~\ref{fig:type}). Looked in relative terms, the advantage of winning the toss is also close to the greatest in multi-day affairs. Only about 60\% of test matches end in a clear decision, the rest end in a draw. Thus, the advantage is closer to 4.5\%. The heftiest raw advantage, however, is in one-day matches (List A and ODIs), approximately 3.3\%. In T20s and T20Is, the advantage is considerably smaller, just about 1.27\%.\footnote{Splitting data by whether the match was domestic or international yields some additional insights. Like \citet{de1998winning}, who based on analysis of data from 427 international one-day matches conclude that `winning the toss at the outset of a match provides no competitive advantage' in one-day international matches, we find that in ODIs teams that win the toss win games at about the same rate as those that lose the toss. For first-class and test matches, the advantage to winning the toss is roughly the same. Meanwhile in Twenty20s, the advantage of winning the toss is greater in international than domestic matches.} And unlike the estimate of advantage for multi-day and one-day affairs, we cannot statistically reject that the possibility that there is no advantage.

\begin{figure}[htbp]
\centering
\caption{Difference in Winning Percentages of Teams that Won the Toss and Teams that Lost the Toss by Day or Day and Night}
\includegraphics[scale=.95]{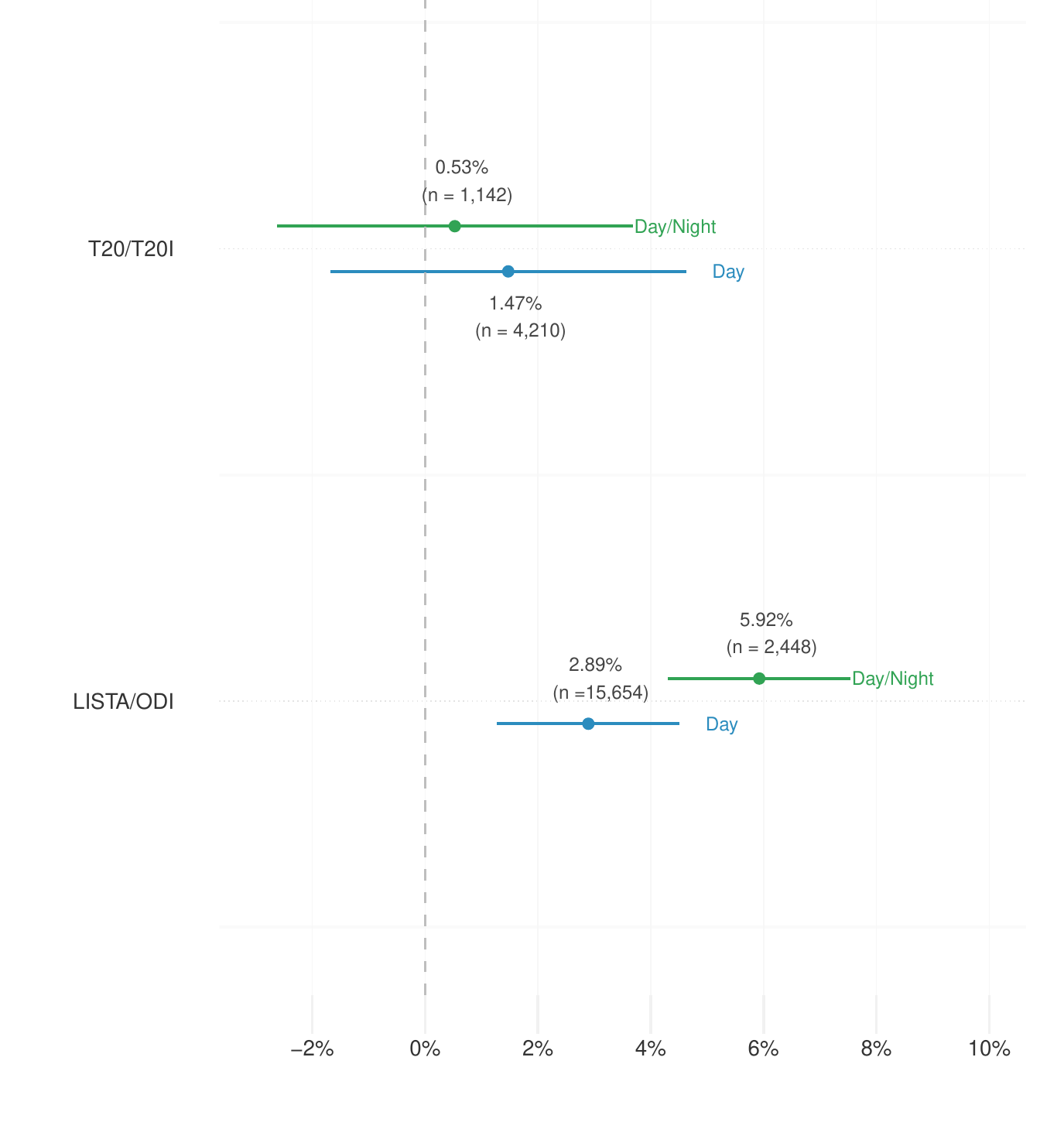}
{\footnotesize \\ Note: Means and 95\% confidence intervals. \emph{n} refers to the number of matches.\par}
\label{fig:dn}
\end{figure}

Type of matches are but one source of variation and theorizing about the advantage granted by the toss. It is often claimed that the toss is more crucial in day and night matches. Due to dew---it is thought to make bowling hard, and the visibility of the white ball is thought to be lower under lights, which makes catching hard---the team that fields second is thought to be at a disadvantage. The conventional wisdom is largely vindicated for one day affairs (see Figure ~\ref{fig:dn}). In one-day matches, the advantage of winning the toss in a day and night match is 5.92\%, whereas the advantage of winning the toss in a one-dayer played during the day is less than half---2.89\%. In Twenty20s---domestic and international---, however, we cannot distinguish between the advantage granted by the toss in day and night matches and day time affairs.

Weather has a large impact on the playing conditions in cricket. For instance, cool overcast weather is thought to aid swing bowling, especially on certain pitches. More generally, the advantage of winning the toss likely varies by weather. However, we do not have data on weather. But, we can proxy it with seasons. In particular, students of the game suspect that the advantage of winning the toss in early English season is especially great. We next assess whether that is so.

\begin{figure}[htbp]
\centering
\caption{Difference in Winning Percentages of Teams that Won the Toss and Teams that Lost the Toss by Month in England}
\includegraphics[scale=1]{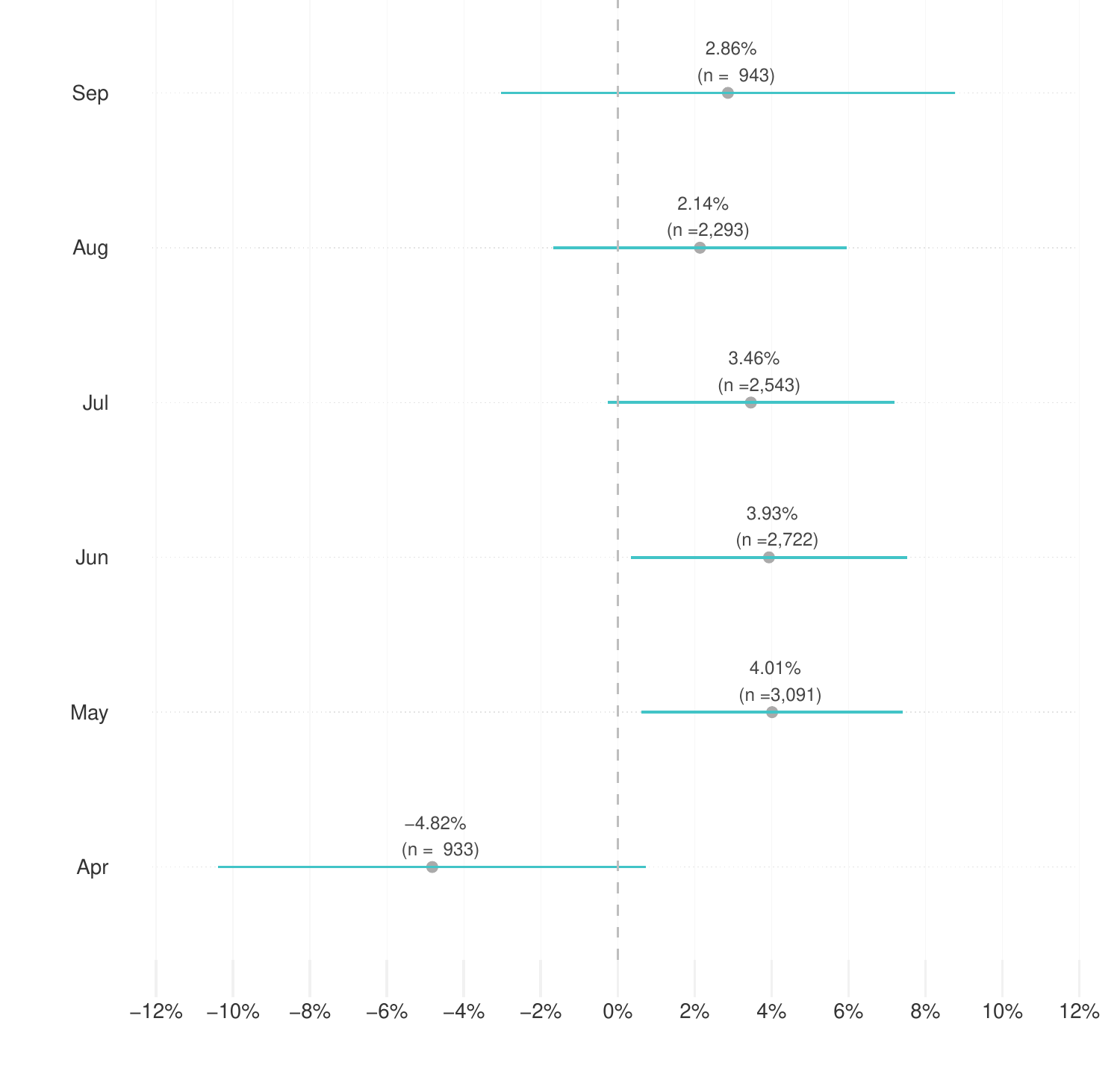}
{\footnotesize \\ Note: Means and 95\% confidence intervals. \emph{n} refers to the number of matches.\par}
\label{fig:season}
\end{figure}

There is some evidence of a seasonal pattern, with advantage of winning the toss somewhat greater in spring and early summer (May and June) than in mid and late summer and early fall (July to September) ~\ref{fig:season}. However, the thing that catches attention is the large disadvantage of winning the toss in April. We don't have a good explanation for the pattern, except for teams choosing badly.  

Aside from affecting the playing conditions, weather affects cricket matches in other, more forceful ways---interrupting, and sometimes ending matches. When a limited over match that is already underway is interrupted by bad weather, and more than a certain amount of time is lost due to the interruption, the match is curtailed and the total that the team batting second must achieve to win is adjusted using a method invented by Duckworth and Lewis \citep[see][]{duckworth1998}.\footnote{Before the Duckworth and Lewis method was adopted, weather affected matches sometimes continued on the next day; an extra day was deliberately left in the schedule for dealing with such eventualities. In cases where the spare day proved inadequate, the match was declared a draw.} 

We can use the random nature of who wins the toss to see if winning percentages of the teams that win the toss are strongly conditioned by whether or not Duckworth-Lewis is used. If the advantage of winning the toss in matches using Duckworth-Lewis is different from matches that don't use it, it suggests that the Duckworth-Lewis method is biased. (For a precise estimate of the bias, ideally, we would want to compare matches using Duckworth-Lewis method with matches held in similar conditions.)

\begin{figure}[htbp]
\centering
\caption{Difference in Winning Percentages of Teams that Won the Toss and Teams that Lost the Toss by Whether or not Duckworth-Lewis was invoked}
\includegraphics[scale=1]{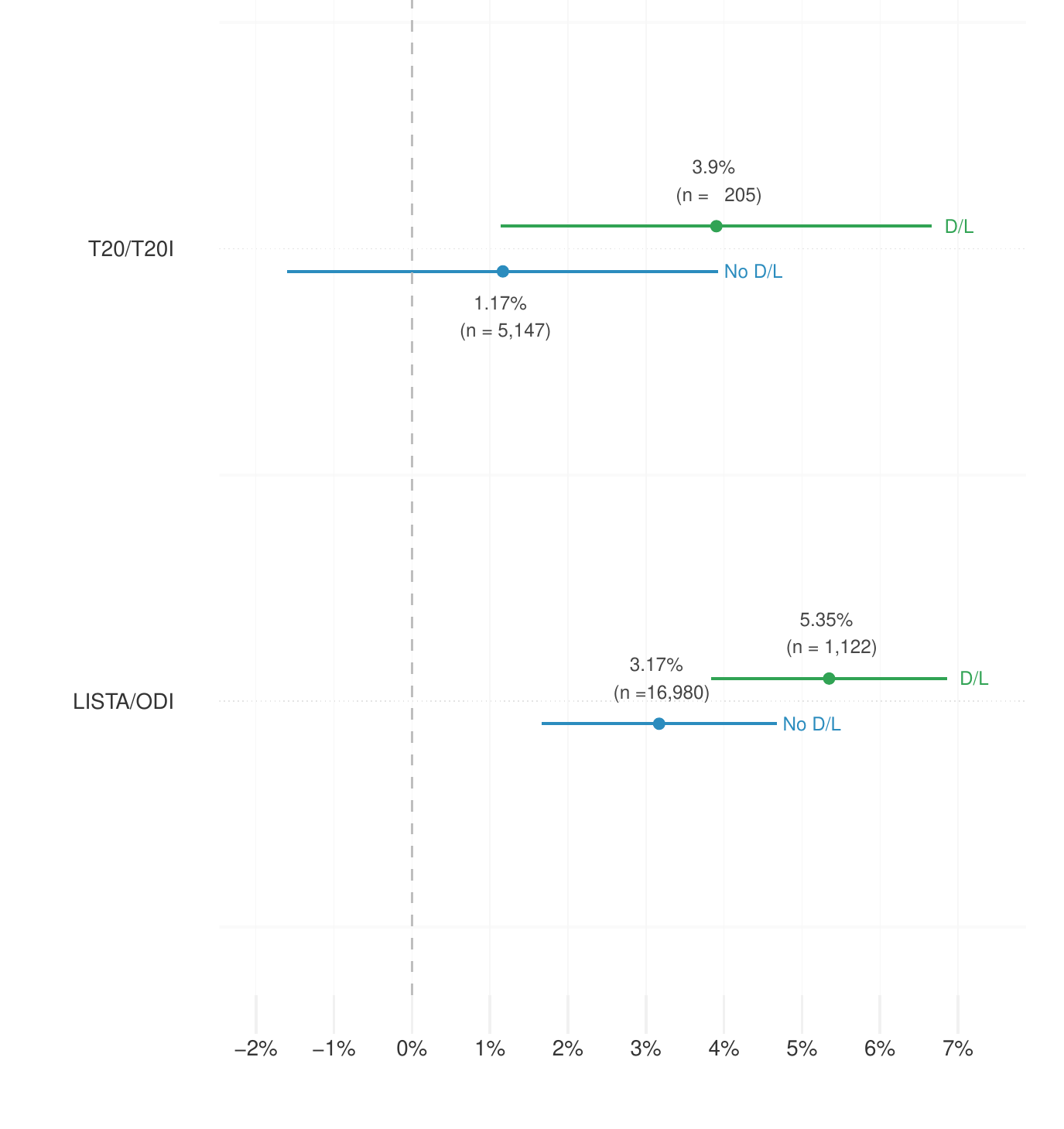}
{\footnotesize \\ Note: Means and 95\% confidence intervals. \emph{n} refers to the number of matches.\par}
\label{fig:dl}
\end{figure}

In both one-day and Twenty20 matches, the advantage of winning the toss in a match where the target is adjusted using Duckworth-Lewis, is considerably greater (see Figure ~\ref{fig:dl}). In one-day matches, the advantage is 5.35\% in matched adjudicated by Duckworth-Lewis and 3.17\% in matches that don't use it. Statistically, the chances that the two numbers are the same is less than 10\%. (And if you make the plausible assumption that winning a toss only improves the chances of winning, the chance that the two numbers are the same is half that.) In Twenty 20s, the advantage of winning the toss shrinks from 3.9\% in matches using Duckworth-Lewis to 1.17\% in matches without it. Once again, the chance that the two numbers are the same is about 10\%. 

Winning the toss ought to matter the most when the difference between the quality of the teams that are playing is the least. Similarly, it is unlikely that winning the toss would change who wins the game when two ill-matched teams are playing. To study the issue, we collected data on team quality. The ICC publishes team ratings for international test and one-day teams each month.\footnote{For details about how the ICC produces these ratings, see \href{http://icc-live.s3.amazonaws.com/cms/media/about_docs/536b1a48c16e5-Reliance\%20ICC\%20ODI\%20Team\%20Rankings\%20FAQs\%202014.pdf}{ICC Rating FAQs}.} Ratings of the men's ODI teams have been published since 1981, and of the test teams, since 1952. Of the entire ranking dataset, that spans 1981--today and 1952--today for ODI and test teams respectively, we only have we have data till 2013.\footnote{The format in which the ICC publishes the ratings changed in 2013. And scraping the latter data posed additional hurdles. We decided that the additional effort wasn't worth the small amount of additional data.} 

Team ratings range from 0 to 143 in our data. For instance, Bangladesh had a rating of 0 in tests for most of 2002 and 2003. And Australia in 2007 twice held a rating of 143. We measure how closely matched the teams by differencing the ranking points of one team from the other. Commercial considerations mean that a majority of the games are played among highly ranked and closely matched teams. Thus, the precision of our estimates is greatest for matches between closely matched teams.

\begin{figure}[htbp]
\centering
\caption{Percentage of Matches Won Minus Matches Lost After Winning the Toss by Difference in Ranks}
\includegraphics[width=1\textwidth]{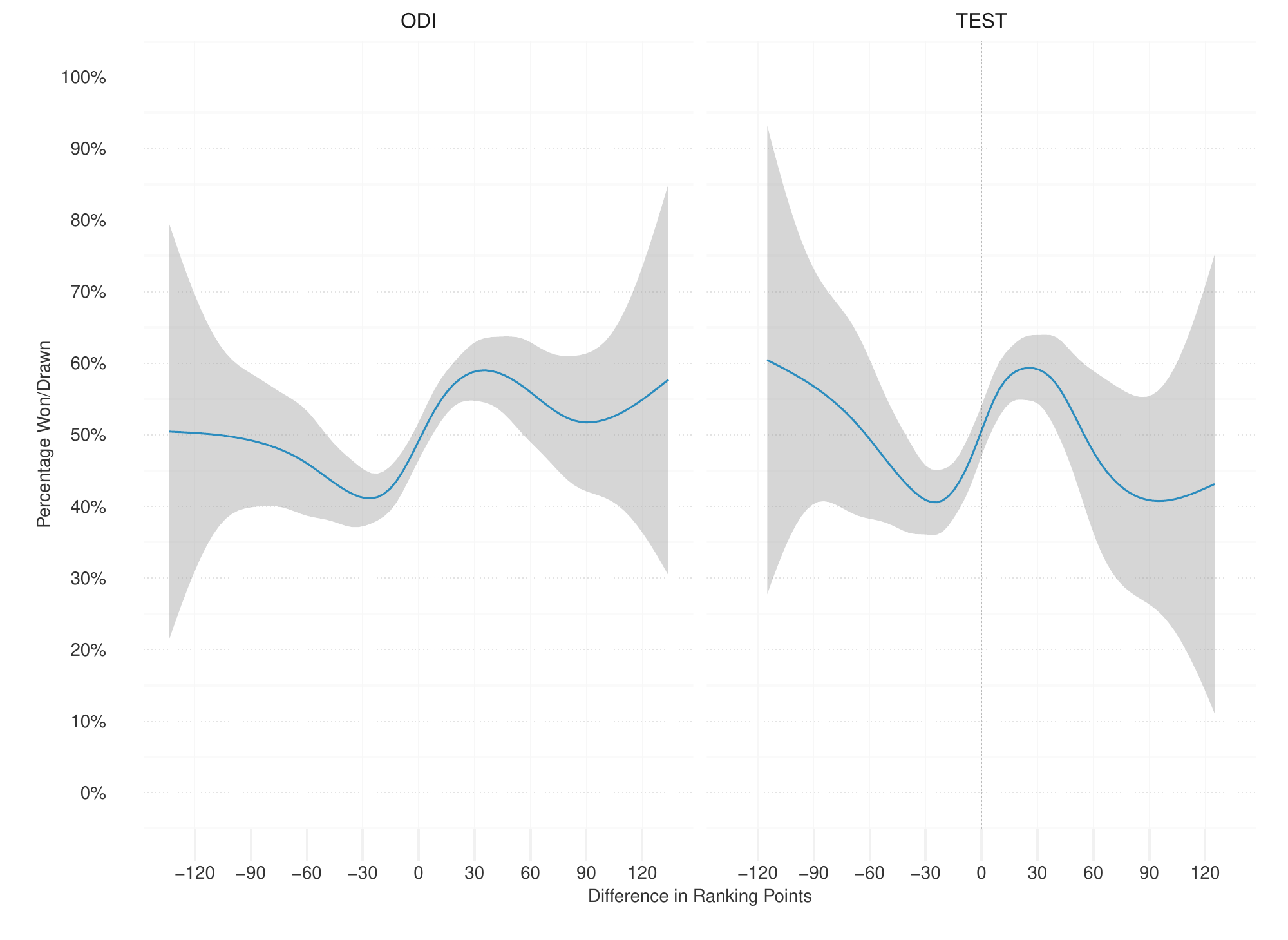}
{\footnotesize Note: Smoothed relationship between difference in ranks and winning probability by whether or not the team won the toss.\par}
\label{fig:ranks}
\end{figure}

The results are expected, but new. As Figure ~\ref{fig:ranks} --- which plots percentage of matches won or drawn by the team that won the toss--- illustrates, there is a sharp curve around 0. When closely matched teams win, winning the toss has a large impact on the probability of winning. 

\begin{figure}[htbp]
\centering
\caption{Difference in Winning Percentages of Teams that Won the Toss and Teams that Lost the Toss by Country}
\includegraphics[scale=1]{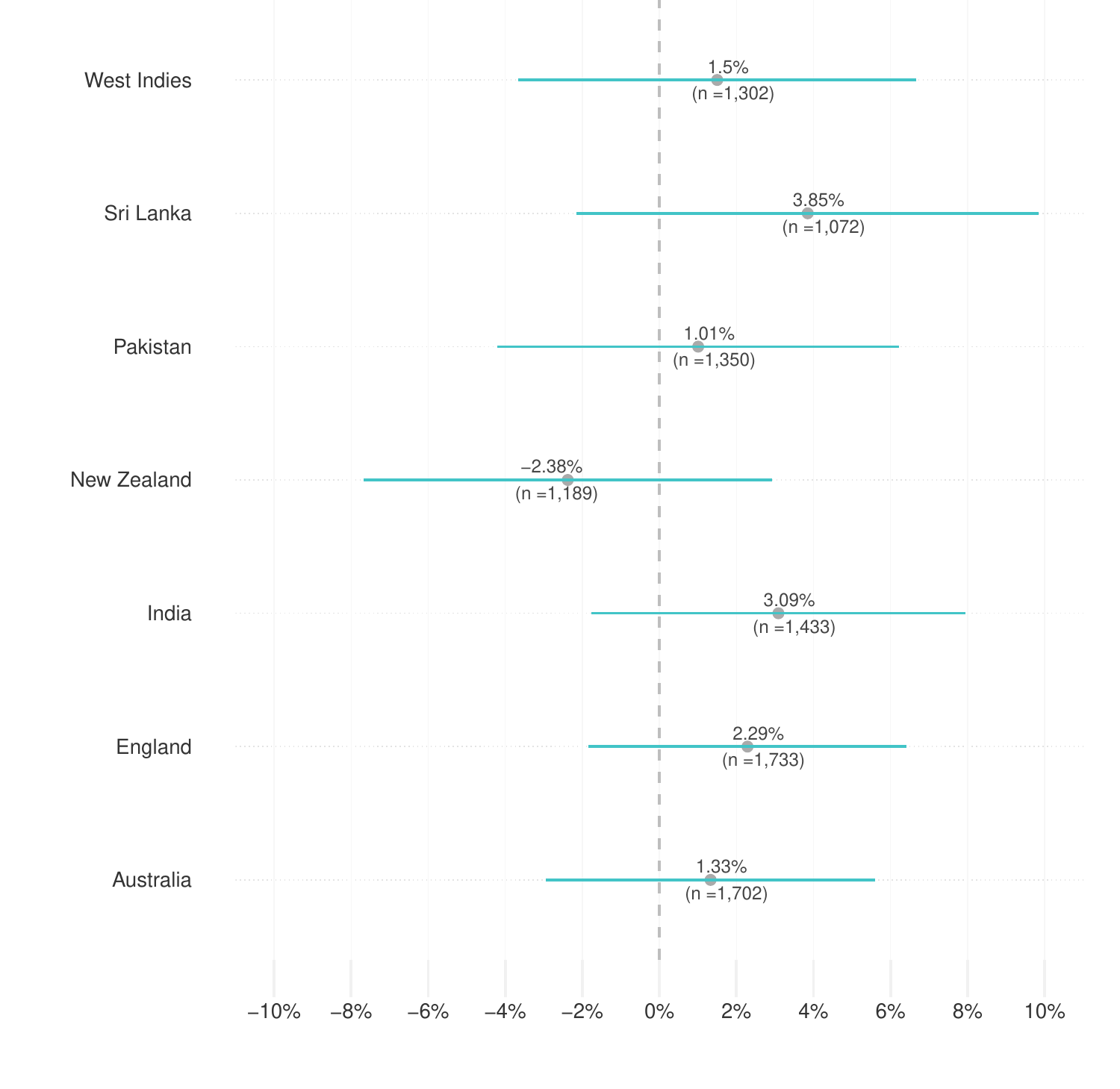}
{\footnotesize \\ Note: Means and 95\% confidence intervals. \emph{n} refers to the number of matches.\par}
\label{fig:country}
\end{figure}

Lastly, we investigate how the advantage varies by country winning the toss in international matches. Are some countries better than others at capitalizing on a toss win? We investigated the question by tallying the advantage by team that wins the toss. As ~\ref{fig:country} illustrates, all of our estimates are imprecise enough that we cannot say with confidence that any of the teams capitalizes on winning the toss. Neither can we discount the possibility that the actual differences across the teams are zero. The most puzzling result is from New Zealand. Like \citet{Saad2015}, data suggest that New Zealand does more poorly in matches where it wins tosses than where it loses it. On the other hand, Sri Lanka and India appear to do especially well, winning 3.85\% and 3.09\% additional matches, respectively, when they win the toss. Australia, Pakistan, and West Indies hover around 1--1.5\%, and England is at 2.29\%. 

Till now, we have focused on assessing the impact of winning the toss on the probability of winning, and how the impact is conditioned by playing conditions, by the type of match, and by the teams involved. Winning the toss, however, likely not only affects the probability of winning, but also the margin of victory. But before we assess the impact of winning the toss on the margin of victory, a short primer. 

In limited over matches, when the team chasing the total falls short, the margin of victory is given in difference in runs. When the team is able to successfully chase the total, the margin of victory is the given by two numbers: number of balls remaining, and the number of wickets in hand. In unlimited overs matches, the metrics for margin of victory differ in two small ways. We don't tally the number of balls remaining when the winning team achieves the target (principally we could). Instead, we note whether or not the winning team had to bat twice---whether or not the team won by an `innings' and additional runs.

Teams that win the toss and the match in first-class and test matches win with more wickets in hand (Mean $= 6.91$) than winning teams that lose the toss (Mean $= 6.64$). In one-dayers and Twenty20s, the teams that win the toss and the match have about the same number of wickets in hand as the teams that lose the toss but win the match (One-Day: Mean$_{\text{Lose Toss}} = 5.57$, Mean$_{\text{Win Toss}} = 5.65$; Twenty20: Mean$_{\text{Lose Toss}} = 6.37$; Mean$_{\text{Win Toss}} = 6.21$). 

In first-class and test matches teams that win the toss and the match also win by few more runs on average (Mean $= 136.48$; Median $=124$) than teams that lose the toss but win the match (Mean $= 133.71$; Median $=122$). Similarly, in Twenty20s, the team that wins the toss wins by a few more runs (Mean $= 37.60$; Median $=28$) than team that loses it (Mean $= 34.50$; Median $=26$). In one-dayers, however, the margin of victory is largely indistinguishable across cases where the winning team wins the toss and where it loses it (Mean$_{\text{Lose Toss}} = 63.95$, Median$_{\text{Lose Toss}} = 51$; Mean$_{\text{Win Toss}} = 63.15$, Median$_{\text{Win Toss}} = 51$). 

Similar patterns hold for balls remaining---teams that win the toss generally win with a few more balls remaining than teams that lose the toss. In one-day matches, Mean$_{\text{Lose Toss}} = 49.87$, Median$_{\text{Lose Toss}} = 28$, Mean$_{\text{Win Toss}} = 53.27$, and Median$_{\text{Win Toss}} = 30$. And in Twenty20s, Mean$_{\text{Lose Toss}} = 18.69$, Median$_{\text{Lose Toss}} = 12$, Mean$_{\text{Win Toss}} = 16.41$, and Median$_{\text{Win Toss}} = 10$. 

And lastly, on number of innings, the teams that win the toss win by an innings about as often as teams that lose the toss (Mean$_{\text{Lose Toss}} = 22.26\%$, Mean$_{\text{Win Toss}} = 22.43\%$).

\section*{Discussion}

The data suggest that winning the toss has a sizable impact on the probability of winning, especially in closely contested games. The data also suggest that the advantage varies considerably and systematically ---in expected ways---, with advantage greater in day and night matches, matches in which Duckworth-Lewis is used to adjust scores, and where the match is played between closely matched teams. In showing so, the data lend credence to, and quantify, the suspicion that many of the cricket fans have long had---that tosses matter. Besides that, the data also help quantify the bias in Duckworth-Lewis method. More generally, the analysis we do here could be replicated elsewhere to assess bias in competing methods, and used to prove that a particular method is better or worse than the Duckworth-Lewis. 

\clearpage
\bibliographystyle{apsr}
\bibliography{luckybib}

\end{document}